\documentclass{elsart}

\usepackage{graphicx}

\usepackage{amssymb}

\begin{document}

\begin{frontmatter}



\title{TPC Performance in Magnetic Fields with GEM and Pad Readout}


\author{D. Karlen\corauthref{cor1}},
\corauth[cor1]{corresponding author}
\ead{karlen@uvic.ca}
\author{ 
P. Poffenberger, 
G. Rosenbaum}
\address{Department of Physics and Astronomy \break
University of Victoria and TRIUMF, Canada}


\begin{abstract}
A novel charged particle tracking device, a high precision time projection
chamber with gas electron multiplier and pad readout, 
is a leading candidate as the central tracker for an
experiment at the International Linear Collider.
To characterize the performance of such a system, a small TPC has been
operated within a magnetic field, measuring cosmic-ray and laser tracks.
Good tracking resolution and two particle separation, sufficient for a
large scale ILC central tracker,
are achieved.
\end{abstract}

\begin{keyword}
time projection chamber \sep gas electron multiplier
\PACS 29.40.Cs \sep 29.40.Gx
\end{keyword}
\end{frontmatter}

\section{Introduction}
\label{intro}
Time projection chambers (TPCs) have been an important part of many
large particle physics experiments
since their initial development in 
the 1970's.~\cite{Nygren:1975nr,Clark:1976kp}
Traditionally, ionization tracks are imaged at the endplates with
wire grids, which provide gas amplification, and pads.
The signals sensed on anode wires and cathode pads are predominantly due to
the motion of positive ions away from the gas amplification region.
When operated in a magnetic field parallel to the drift field,
the momentum resolution of the device is limited by non-zero 
{\bf E}$\times${\bf B}~\cite{Hargrove:1982yy}
in the vicinity of the wire grids.
The multi-track resolution is limited by the 
wide pad response function and the slow drift velocity of the positive ions.

A time projection chamber is a leading candidate to be
the main tracker for an
International Linear Collider experiment.~\cite{Behnke:2001qq}
In order to improve the momentum and multi-track resolutions, 
as required by the physics objectives, it has been proposed that the
wire grids be replaced by a micropattern gas avalanche detector, such
as a gas electron multiplier (GEM)~\cite{Sauli:1997qp}
or micromegas~\cite{Giomataris:1995fq}.
For these types of devices, {\bf E}$\times${\bf B} can be negligibly small.
Furthermore, since the pad signals are due to the motion of electrons as they
approach and arrive on the readout pads, the spatial extent of the signals can be much 
narrower and their risetimes much faster, thereby improving the multi-track resolution.

In fact, the signals can be so narrow as to
present a challenge for conventional pad readout in a large detector.
The outer radius of a linear collider experiment
main tracker is envisaged to be approximately 2~m and
to keep electronics costs reasonable, 
the readout pads need to be no smaller than about 2~mm~$\times$~6~mm.~\cite{Behnke:2001qq}
A strong magnetic field of 4~T is being considered to reach the momentum resolution goal.
In this magnetic field,
gases with fast drift velocity at low drift fields can have transverse
diffusion contants of about 30~$\mu$m/$\sqrt{\mathrm{cm}}$.
To achieve optimal transverse resolution with 2~mm wide
pads, a mechanism to defocus the drifting electron
charge cloud after amplification is required, so that
the signals are sampled by at least 2 pads per row.

One way to defocus the charge cloud is to use
gas diffusion between the GEM foils and the readout pads.
In these regions the electric field can be much larger
and by selecting an appropriate gas, the diffusion constant can
be much larger there than in the drift region.
This is the method that is investigated in the work reported here.
Ideally, the defocusing should be sufficient so that the standard deviation
of the charge clouds should be at least ${1\over4}-{1\over3}$ of the
pad width, as explained in appendix~\ref{chargeSharing}.
An alternative defocusing method, 
particularly well suited for micromegas devices, is to
place a resistive foil above the pads.~\cite{Dixit:2003qg}

The remaining sections of the paper describe studies with
a TPC consisting of a drift volume coupled to a double GEM structure 
and conventional pads.
In section~\ref{system}, the design and general properties of the
TPC and the laser delivery system are described.
The data samples collected with this device are summarized
in section~\ref{data}, and the general properties of the
data are shown.
Section~\ref{simul} describes a program used to
simulate cosmic and laser tracks in
the TPC, which was used to develop
and test reconstruction algorithms.
In section~\ref{trackFitting} the methods employed to fit
tracks of ionization in the TPC are described in detail.
Section~\ref{results} 
shows the performance of the TPC,
as measured under a variety of operating conditions,
and compares these to results from simulations.

\section{TPC and laser system}
\label{system}
After gaining experience operating a 15~cm
drift length TPC with GEM readout without
magnetic fields~\cite{Carnegie:2004cu}, 
a new TPC was designed with a 30~cm drift length,
specifically to be deployed in magnets available for use at the 
TRIUMF and DESY laboratories in 2003.
Following successful cosmic ray tests in the magnets that year, a laser
delivery system was constructed for further tests with the TPC in
the DESY magnet in 2004.

\subsection{TPC design}
\label{design}
\begin{figure}
 \begin{center}
  \includegraphics[height=2in,clip]{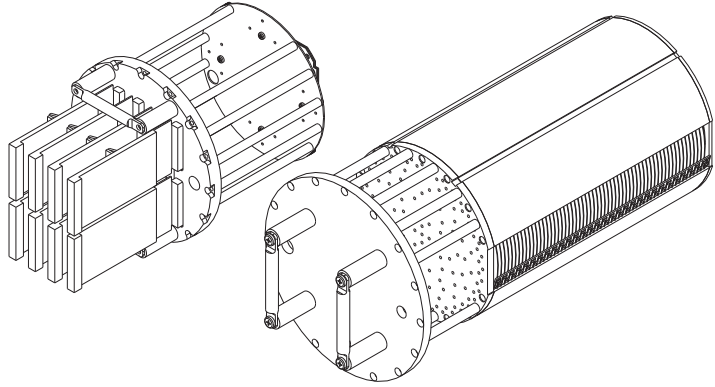}
 \end{center}
 \caption{
  Design drawings of the readout module (left) and drift module (right). 
  The two modules
  are inserted into either ends of a cylindrical acrylic tube to make a 
  closed gas volume.
  The readout module shows the 8 front end cards of the electronics readout
  system inserted.
  \label{fig:tpcpieces}
 }
\vskip 1cm
\end{figure}

The TPC is a cylindrical design, constructed primarily with acrylic.
The drift length is 30~cm with a field cage of brass hoops with a pitch of
5~mm held in slots in an acrylic holder.
The potentials of the hoops are provided by a chain of 10~M$\Omega$ surface
mount resistors in the gas volume.
The cathode endpiece is a circular piece of G10 with copper cladding.
The other end of the drift volume is terminated by a circular endpiece with a
10~cm square hole, equal to the size of the GEM foils.
The endpiece has copper cladding, and narrow wires strung across the
hole, to provide a well defined termination of the electric field.
The drift volume section of the TPC is inserted into an outer
acrylic cylinder, with an outer diameter of 8.75 in.
A readout module, holding a pad array constructed on
a printed circuit board and two GEM foils, is
inserted into the cylinder on the other end, to make a closed
gas volume.
The readout and drift modules design drawings are shown in Fig.~\ref{fig:tpcpieces}.

\subsection{Drift field and GEM operation}
\label{hv}
\begin{figure}
 \begin{center}
  \includegraphics[height=2in,clip]{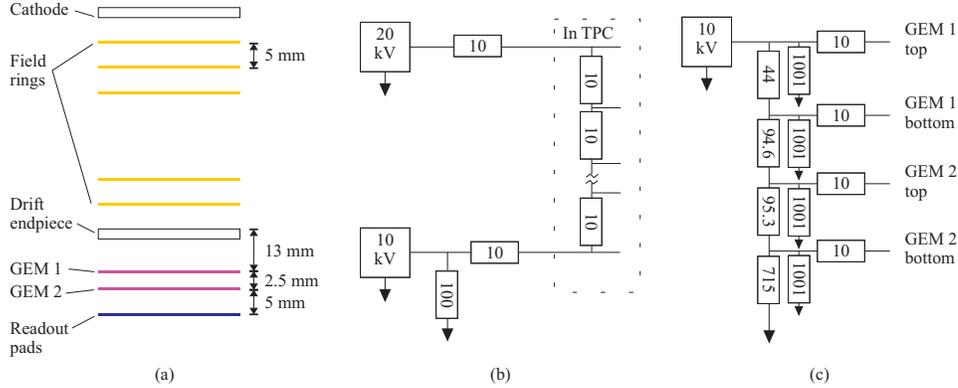}
 \end{center}
 \caption{
  The configuration of the high voltage system is illustrated. (a) The orientation of the
  components. (b) The high voltage for the drift volume is controlled by two power
  supplies (maximum ratings shown). The rectangles show resistances in M$\Omega$. The
  100~M$\Omega$ drain resistance is provided so that different drift fields can be
  applied without the lower supply 
  sinking current. (c) The high voltage distribution
  for the GEM foils. The 1001 M$\Omega$ resistances indicate the
  1000:1 dividers for monitoring the voltages. Not shown are
  the isolated voltmeters across shunts to monitor
  the GEM currents.
  \label{fig:hvsystem}
 }
\vskip 1cm
\end{figure}

A schematic diagram of the high voltage configuration is shown in 
Fig.~\ref{fig:hvsystem}.
All resistors shown are installed in a HV distribution box, except for the field cage
network, which are surface mount resistors located in the TPC volume.
The HV distribution box also contained 100~k$\Omega$ shunt resistors on one line
to each GEM, to which isolated digital voltmeters are attached.
This allowed for current monitoring at the level of 1~nA on each GEM.

During data taking,
electric fields in the drift volume were 90-250~V/cm, chosen to maximize
the drift velocity, and
the field between the drift endpiece and GEM~1 was 10~V/cm larger.
The transfer field between the two GEMs was about 2.5~kV/cm, and the
induction field to the readout pads was about 3.5~kV/cm.
The potential across each GEM was 370-380~V, providing
an effective gas gain of the system in the range 5-8$\times 10^3$.

\subsection{Readout electronics}
\label{readout}
An array of rectangular readout pads, roughly 10~mm$^2$ in area are laid out on a
printed circuit board, arranged in rows of 31 or 32 pads across.
The pads are at nominal ground potential, and each row is connected to
a front end card developed for the STAR TPC~\cite{Anderson:2002qr}.
The card contains charge sensitive preamplifiers-shapers and 
when a trigger is received, the amplified signals are sampled and stored into
a 512 time bin switched capacitor array.
The card was modified to increase the pedestal to allow negative
going pulses to be sampled with a large dynamic range.
The signals are digitized asynchronously and 10~bits per channel
per 50~ns time bin are stored in the raw data files.
For cosmic ray studies, trigger signals are formed
from the coincidence of scintillator paddles placed above and below the TPC.
For laser studies, trigger signals are provided by a photodiode.

\subsection{Laser system}
\label{laser}
\begin{figure}
 \begin{center}
  \includegraphics[height=2.5in,clip]{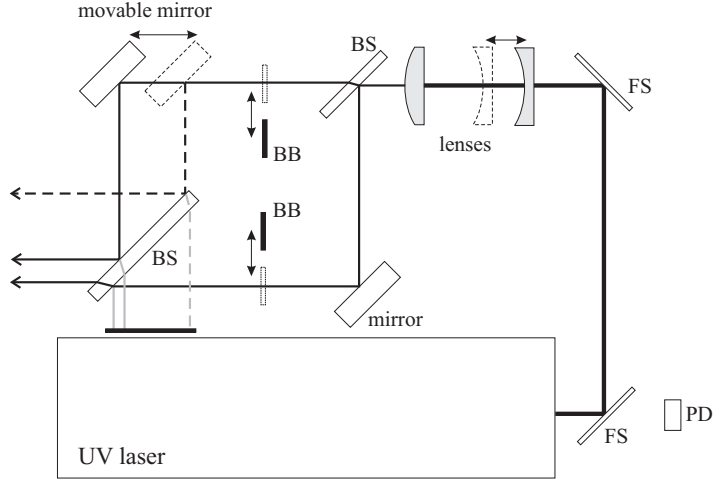}
 \end{center}
 \caption{
A schematic drawing (not to scale) of the laser system in plan view.
(PD) is a photodiode used for triggering the readout electronics.
(FS) are fused silica windows, used to reflect the beam and reduce
its intensity.
(BS) are beam splitters.
(BB) are movable beam blockers to block either beam.
The movable mirror allows the horizontal separation of the beams to be adjusted.
The beams are transported below the TPC, and are reflected up through the TPC
drift volume between the field rings, by a movable mirror, and
optionally by a movable beam splitter.
  \label{fig:laser}
 }
\vskip 1cm
\end{figure}

In order to perform controlled studies of the GEM TPC, a laser
delivery system was constructed for use in the DESY magnet system
with an ultraviolet (266~nm wavelength) Nd:YAG laser.
The system is able to provide a single beam or a pair of beams
to the TPC perpendicular to the drift direction, at any
location along the drift distance, under remote control.
The TPC outer acrylic tube was fitted with long quartz windows
to accept the laser light.
A schematic drawing is shown in Fig.~\ref{fig:laser}.

The beam from the laser is reflected by two fused silica slides (each with the back face
sandblasted to reduce ghost reflections) to reduce the beam intensity
and focused by a $F=-30$~mm
plano-concave lens followed by a $F=100$~mm plano-convex lens.
The beam is split, with the reflected beam providing a fixed direction,
and the through going beam reflected off a movable mirror to
provide a second beam parallel to the fixed beam.
The separation of the two beams can be adjusted to between 0 and 12~mm.
Beam blockers can move into or out of either beam to provide a single beam
to the TPC.

When the TPC is placed in the laser delivery system holder, a movable splitter
and mirror are underneath the TPC to reflect the beam
into the TPC volume at two drift distances simultaneously.
The splitter can be rotated out of the beam so that the light enters at
only one drift distance.
A steering mirror was also in place to allow a transverse
separation of beams that enter at two drift distances.

Remote control was provided through the use of servos and stepper motors
controlled via the serial and parallel ports of the data acquisition computer.
Operation under remote control was necessary for personal safety considerations
regarding the strong magnetic field and the ultraviolet laser.

\section{Data samples and characteristics}
\label{data}
In 2003, cosmic ray data were collected at a variety of magnetic field strengths
for two different gas mixtures:
P5 (Ar:CH$_4$ 95:5) and a mixture referred to as TDR-gas (Ar:CH$_4$:CO$_2$ 95:3:2).
The pad pitch was 2~mm~$\times$~7~mm.
Initial tests were performed in the 1~T warm magnet at TRIUMF followed by
a longer data taking period in the superconducting magnet at DESY.
Two of the electronics cards were faulty during much of the data taking run,
but 6~rows were found to provide sufficient information to characterize
the tracking resolution performance.

In 2004, cosmic ray and laser calibration data were collected primarily at
a magnetic field of 4~T at DESY, with the same two gas mixtures.
Two readout pad boards were used: the original with 2~mm~$\times$~7~mm pads, and
a new board with 1.2~mm~$\times$~7~mm pads.
The study with the narrower pads was performed because it was found in the 2003 run that
neither gas provides adequate defocusing between the GEM foils for the
2~mm wide pads at high magnetic fields.
Gas mixtures with higher concentrations of methane, such as P10, could provide the
required defocusing at 4~T, but this gas was not permitted in the DESY 
magnet test area.
The initial data taking in 2004 was with the wider pads, after which the 
TPC was briefly opened to insert the narrow pad plane.
The data taking started with P5 gas when it was found that the drift velocity
at 90 V/cm, where it is expected to be maximum, was found to be 
much lower than expected and increasing with time.
By operating at 160~V/cm, the drift velocity was maximised and found to be
stable.
After the data taking period was completed, it was found that long lengths of
the gaslines were copper that had been previously exposed to atmosphere, and a gas 
analyzer showed that under these conditions, significant amounts of water 
remain in the system for a few weeks after starting the gas flow.
The drift velocity and diffusion of TDR gas is less sensitive to water.

The 2003 data had several faulty electronics channels, unlike the 2004 data,
and since the laser system was only incorporated for the 2004 data set, the focus of the
data analysis reported in this paper is on the 2004 data samples, summarized in 
table~\ref{tab:datasamples}.

\begin{table}
\begin{center}
\caption{
List of cosmic data samples recorded in 2004, in order of data taking.
The name assigned to the each configuration is referred to in
subsequent tables. The first data set likely contained a large
concentration of water.
\vskip 5mm
}
\begin{tabular}{|l|c|c|c|c|}
\hline
Name    & Gas    & B field & Pad pitch & Drift field \\[-8pt]
        &        & [T]     &   [mm]    &  [V/cm]     \\
\hline
p5B4w   & ``P5'' & 4.0     & 2.0       & 160         \\
\hline
tdrB4w  & TDR    & 4.0     & 2.0       & 230         \\
\hline
p5B4n   & P5     & 4.0     & 1.2       & 90          \\
\hline
tdrB1n  & TDR    & 1.0     & 1.2       & 230         \\
\hline
tdrB0n  & TDR    & 0.0     & 1.2       & 230         \\
\hline
tdrB4n  & TDR    & 4.0     & 1.2       & 230         \\
\hline
\end{tabular}
\label{tab:datasamples}
\end{center}
\vskip 1cm
\end{table}

\begin{figure}
\begin{center}
\includegraphics[width=0.9\textwidth,clip]{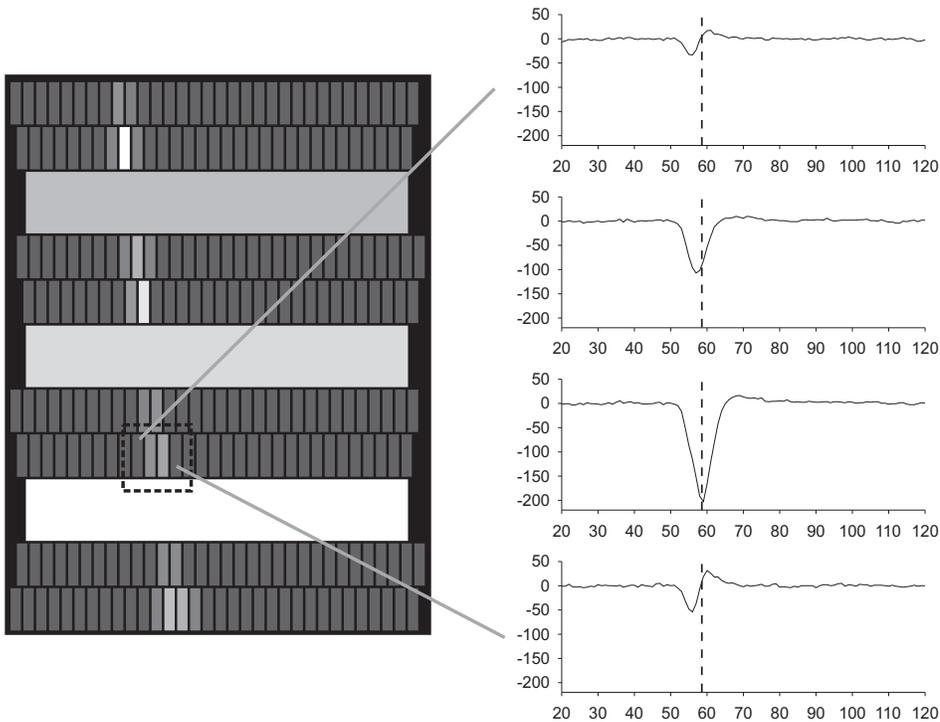}
\caption{A typical cosmic ray track observed with the wide pad layout
at 4T with P5 gas.
The brightness of the pads indicates the amount of charged collected.
The three very wide pads are used during data taking
for selecting events that contain a track in the active area.
The pulses for 4 neighbouring pads are shown on the right.} \label{fig:event}
\end{center}
\vskip 1cm
\end{figure}

A typical cosmic ray event is illustrated in Fig.~\ref{fig:event}, along with
representative pulses on some of the pads.
Pads near the center of the track
have signals due to the collected charge that are
significantly different in shape than pads further away, which
have only signals induced by the motion of electrons in the induction gap.
Due to the shaping of the readout electronics, the induced pulses are
transformed into bipolar pulses.

To estimate the charge collected by pads, clusters of signals in a row
are searched for, and the time bin with the largest sum of signals in neighbouring
pads is taken to be the arrival time of the center of the electron cloud.
For each pad in the row, 
the sum of the pulse height in that time bin and the three previous and
three following time bins is formed to estimate the collected charge.
Studies show that the induced signals roughly cancel out with this
algorithm.

\begin{figure}
\centering
\includegraphics[width=120mm,clip]{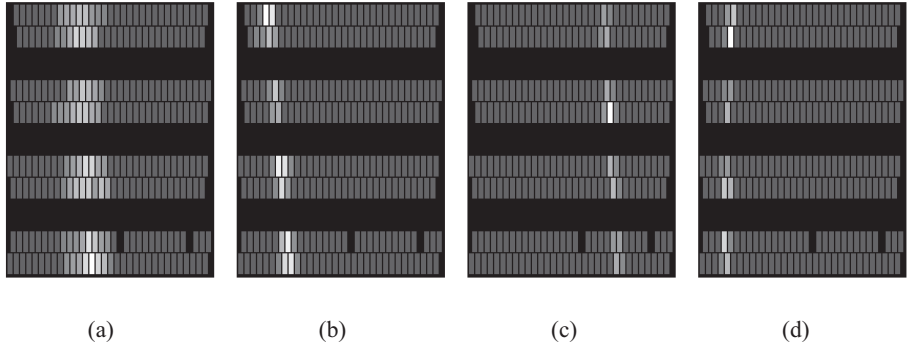}
\caption{Typical cosmic ray tracks observed with the wide pad layout
at different magnetic fields with P5 gas. 
For all events the drift distance is approximately 25~cm.
The brightness of the pads indicates the amount of charged collected.
The black pads are disabled due to electronics problems.
(a)~0~T, (b)~0.9~T, (c)~2.5~T, (d)~5.3~T.
} \label{fig:manyevents}
\vskip 1cm
\end{figure}

Figure~\ref{fig:manyevents} shows cosmic ray events at the same drift
distance recorded with different magnetic fields, and the effect of
reduced diffusion in the drift volume due to the magnetic field is apparent.

\section{Simulation}
\label{simul}
A simulation package was created in order to develop reconstruction
methods and better understand the various data distributions.~\cite{Karlen:jtpc}
Ionization electrons can be inserted into a gas volume either along a 
straight trajectory using
a parameterization of ionization fluctuations (to simulate muon tracks) or
Poisson fluctuations (to simulate
laser tracks), or along a realistic trajectory by using the energy loss of 
particles traversing
the gas, as calculated by the GEANT3~\cite{Brun:1987ma} program.
For most of the simulation studies presented here,
the latter approach was used with GEANT3 description of the DESY test setup and
a cosmic muon generator~\cite{McPherson:cosmic}, 
in order to reproduce the cosmic muon acceptance.

The ionization electrons are transported in the gas towards the TPC endplate according to
a specified drift velocity and diffused according to given transverse and
longitudinal diffusion constants.
The electrons enter the GEM foil holes, with a given collection efficiency,
are amplified according to either a Poisson or an exponential gain function,
and exit the holes with a given extraction efficiency.
Between the foils and between the second foil and the pad plane they diffuse,
with generally larger diffusion constants.
The electrons are collected by an array of pads, and the pad signals are
shaped, digitized, and saved in simulated data files that can be analyzed by the
same software that analyzes the real data.

For the simulated samples shown in this paper, 
the drift velocity and diffusion
constants were set to be the values deduced from the data or from 
Magboltz~\cite{Biagi:magboltz}
simulations, the collection and extraction efficiencies set to 1, and an
exponential gain function assumed for each GEM foil.

\section{Data Analysis}
\label{trackFitting}
As part of the simulation package~\cite{Karlen:jtpc}, methods were
developed to find and reconstruct ionization tracks.
A simple track finder uses the largest clusters from the outermost rows
to define straight trajectories as a starting point for the track fitter.
Track fitting is
performed separately in the readout pad plane and 
in the vertical-drift plane.
The spatial resolution is most critical in the
readout pad plane, since it directly determines the
momentum resolution of a TPC.
To achieve the ultimate resolution, a special track fitting
algorithm was developed, and is described in detail below.
No channel by channel calibration constants or 
empirical corrections were applied in
the analysis.

\subsection{Tracking in the pad readout plane}
\label{rphi}
To accurately
reconstruct the ionization tracks from the measurements of charge collected by the
readout pads, a maximum likelihood track fitting algorithm was developed.
This takes into account the non-linear charge sharing function, 
which is necessary to achieve good resolution when
the pads are as wide as the distribution of electrons arriving on the pads.
An alternative method, using 
a simple linear centroid determination of the track coordinates as it crosses each row
of pads, was found to perform poorly under these conditions.

A right handed coordinate system is used to define a set of track parameters,
$x_0$, $\phi_0$, $1/r$, and $\sigma$, that describe ionization tracks projected into
the pad readout plane.
The $x-y$ plane is parallel to the pad plane and
perpendicular to the drift direction, the $+y$ direction is
the vertical direction, and the $+z$ direction is opposite to the drift direction.
The choice of the coordinate origin, $(x=0,y=0,z=0)$,
is chosen to be at the centre of the readout pad array.
The location of the track as it crosses the $x$ axis is $x_0$ and its local azimuthal
angle at that location is $\phi_0$.
Vertical tracks are assigned an angle $\phi_0=0$, and positive angles correspond to
right handed rotations of the track about the $z$ direction.
To account for the helical trajectories of charged particles in an axial magnetic field,
the inverse radius of curvature, $1/r$, is included, with positive
values assigned to those whose centre of curvature is displaced
with respect to the track in the $+x$ direction.
The final track parameter, $\sigma$, is the standard deviation of the
charge distribution about the trajectory as measured at the pads.

The maximum likelihood track fit is described in detail in appendix~\ref{xyTrackFit}.
A relatively simple model is used to describe the expected charge distribution across
the readout pads, given the track parameters defined above.
As usual, estimates for the track parameters for each event are made by
maximizing the likelihood function.
Provided enough rows of data are recorded and good starting values for the parameters
are used, the maximum of the likelihood function is converged upon quickly using
standard numerical techniques.

The second derivatives of the likelihood function can be used to estimate the covariance
of the track parameters.
The error matrix elements estimated in this fashion are sensitive to
the model assumptions and its parameters, and therefore the resolution are
estimated directly from the data, by measuring the standard deviation of residuals from
repeated measurements.

One effect that causes the errors estimated by the fit to be underestimated is the
non-uniform ionization that occurs within a row.
Except for tracks that are parallel to pad boundaries, this increases the variance of the
probability density function for charge deposition on pads near the track.
The resulting degradation of track resolution for non-vertical tracks is
referred to as the ``track angle effect''.
Another important effect that degrades the resolution is 
the variance of the single-electron response
of the GEM, which reduces the
statistical power of the original ionization electrons.

One effect that causes the errors to be overestimated is the model
assumption that the diffusion within and after the GEM structure is small.
The observed distribution of charge across the pads is used to estimate the diffusion
of the ionization electrons, whereas in reality, the distribution is due both to the
diffusion in the drift volume and the diffusion of the amplified electrons in
the GEM structure.
Since the multinomial distribution that is used refers to the original 
ionization electrons, the
estimated uncertainties from the track fit will be overestimated.

\subsection{Two track fitting in the pad plane}
\label{twoTrackFitting}
If two ionization tracks are near enough, some pads will collect charge from
both tracks, and the tracking resolution for each track will degrade.
To analyze data with two nearby tracks, 
the maximum likelihood track fit was modified to include the
charge from two tracks.
To allow for non-uniform ionization along the track direction, 
the ratio of the charges sampled 
from the two tracks is treated as an additional nuisance parameter for each row.
In other words, the relative amplitudes of the pulses from each track 
are not assumed to be
one half, or some other fixed value, but instead the relative amplitudes for each row are
varied in the maximization of the likelihood.
In these studies, the width of the charge distribution for each track, $\sigma$,
is fixed.

\subsection{Tracking in the vertical-drift plane}
\label{rz}
In the vertical-drift plane, a more traditional approach is applied.
For each row, the drift time is determined and converted to drift distance.
The set of points defined by the pad row centres and the drift distances
are fit to a straight line, by minimizing the squares of the deviations.

\subsection{Selection criteria}
\label{criteria}
A study of noise present in all channels identified three bad channels 
for the entire 2004 data run
and a few single bad channels for subsets of the data run.
The information from these pads is ignored in the analyses presented here.

Events are selected containing a cluster of signals in each row, 
with the threshold set to be 
typically over 99\% efficient per row for cosmic ray tracks.
Multiple track events are removed by requiring no more than one cluster per row.
The reconstructed track is required to have its $x_0$ coordinate well within
the geometric acceptance of the pads, and the azimuthal angle is required to be
$|\phi_0| < 0.1$.
Likewise, the $z_0$ coordinate is required to be well within the drift volume, and 
the dip angle $\lambda$ is required satisfy, $|\tan\lambda|<0.6$.
A small number of events are removed by excluding those with an 
anomalously large $\sigma$ 
which arise when a very large ionization, due to a delta ray, cause large induced signals
on a large number of pads in a row.
Finally a few events are removed by requiring that the track fit converge properly, and
the estimated errors on $x_0$ and $\sigma$ be within a reasonable
range, to ensure that the fit results are sensible.

\section{Results}
\label{results}

\subsection{General properties}
\label{generalProperties}
The drift velocity is quickly determined with the laser system by measuring the
arrival times of laser tracks at known drift distances.
This was also used to ensure that the drift velocity was relatively constant
throughout a data run.
The results are shown in table~\ref{tab:gasprop}.

\begin{table}
\begin{center}
\caption{
Measurements of the drift velocity, diffusion, and defocusing, 
for the 2004 cosmic data sets compared to expected values.
The data set names are defined in table~\ref{tab:datasamples}.
The columns labelled ``$v_d$ sim'' and ``$D$ sim'' shows the results from a
Magboltz~\cite{Biagi:magboltz} simulation, whereas the final column shows the
results from the full TPC simulation, using Magboltz values as the input parameters for
the gas properties.
\vskip 5mm
}
\begin{tabular}{|l|c|c|c|c|c|c|}
\hline
Data    & $v_d$       & $v_d$ sim   & $D$     & $D$ sim  & $\sigma_0$ & $\sigma_0$ sim \\[-8pt]
        & [cm/$\mu$s] & [cm/$\mu$s] & [$\mu$m/$\sqrt\mathrm{cm}]$ 
                          & [$\mu$m/$\sqrt\mathrm{cm}$]  & [$\mu$m]   & [$\mu$m]       \\
\hline
p5B4w   & $3.84\pm0.08$ & 3.64 & $76\pm5$   & $67\pm1$   & $429\pm2$  & $350\pm2$  \\
\hline
p5B4n   & $3.85\pm0.04$ & 4.14 & $34\pm5$   & $43\pm1$   & $382\pm1$  & $369\pm1$  \\
\hline
tdrB4w  & $4.51\pm0.05$ & 4.52 & $71\pm10$  & $69\pm1$   & $367\pm4$  & $262\pm1$  \\
\hline
tdrB4n  & $4.54\pm0.06$ & 4.52 & $70\pm5$   & $69\pm1$   & $319\pm3$  & $255\pm1$  \\
\hline
tdrB1n  & $4.66\pm0.06$ & 4.52 & $205\pm10$ & $206\pm2$  & $509\pm2$  & $289\pm2$  \\
\hline
tdrB0n  & $4.68\pm0.06$ & 4.52 & $348\pm20$ & $468\pm10$ & $918\pm15$ & $580\pm1$ \\
\hline
\end{tabular}
\label{tab:gasprop}
\end{center}
\vskip 1cm
\end{table}

\begin{figure}
\centering
\includegraphics[width=66mm,clip]{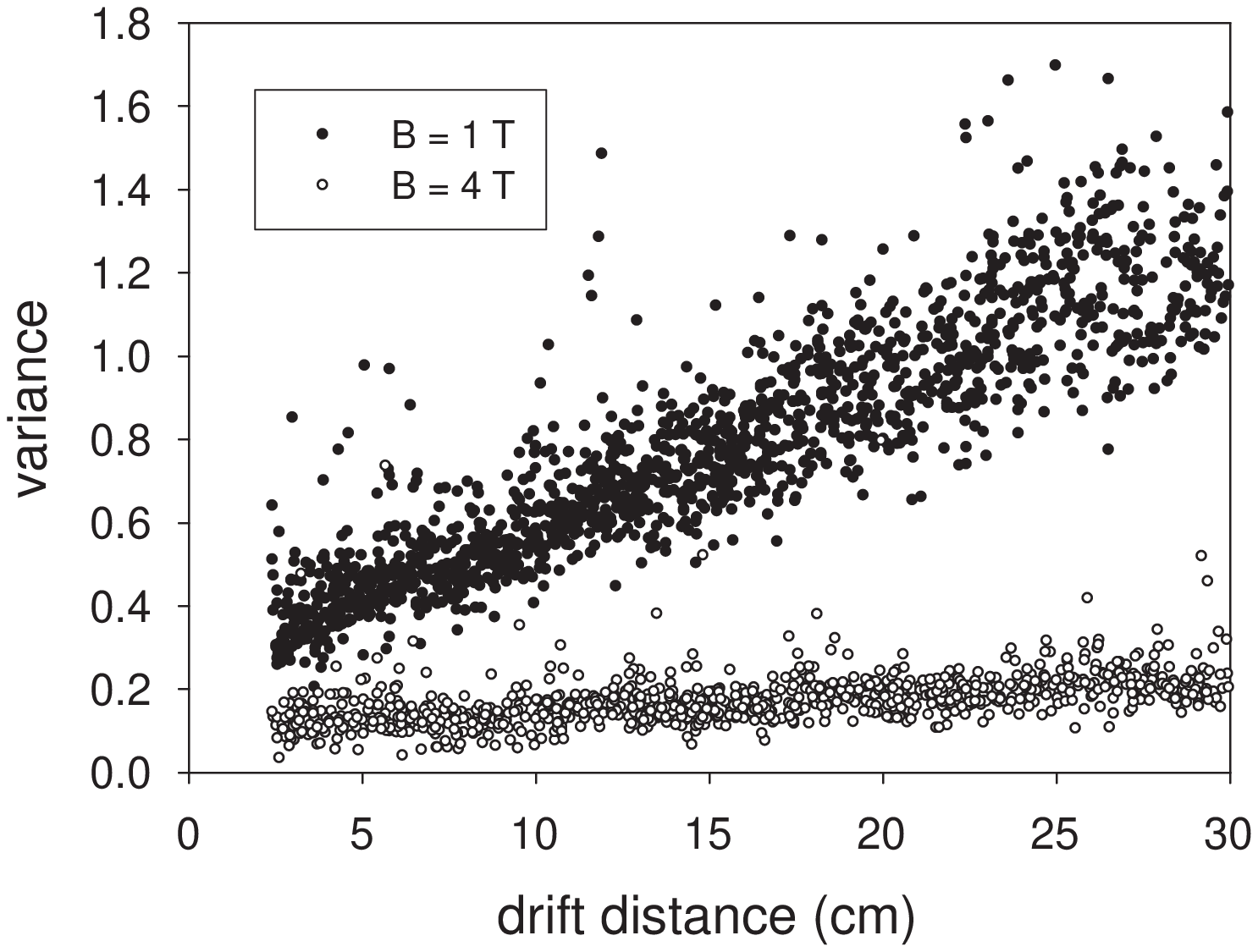} \ \ \ 
\includegraphics[width=66mm,clip]{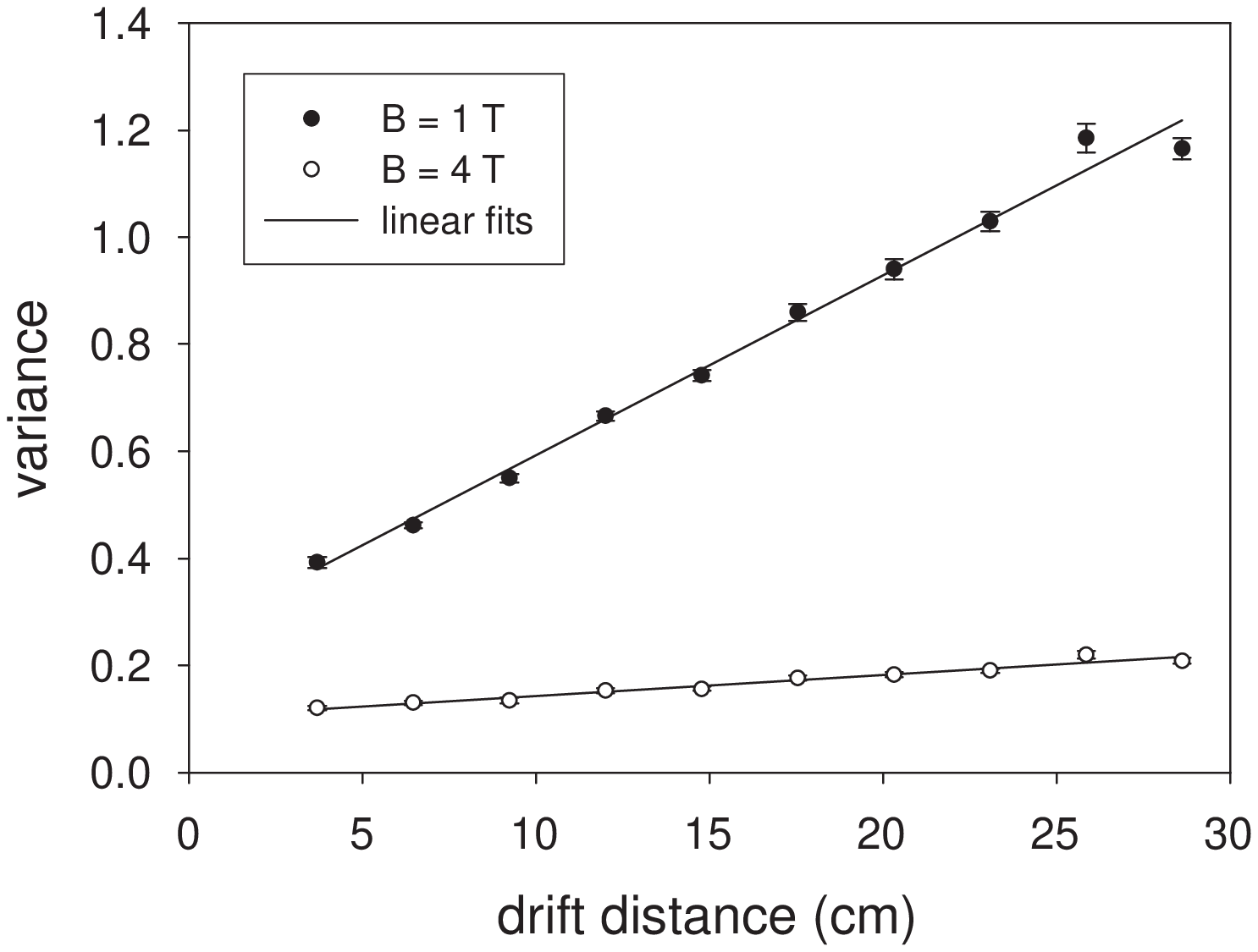}
\caption{
On the left, the points show the variance (ie. $\sigma^2$ in mm$^2$) 
and the $z_0$ for events
in TDR gas at $B=1$ and 4~T.
The right plot shows this data divided into 10 bins of drift distance, 
with the mean and error in the
mean shown, along with linear fits to the data.
} \label{fig:diffusion}
\vskip 1cm
\end{figure}

The diffusion properties of the gas, both in the drift volume and in the amplification
section determine, to a large extent, the performance of the TPC.
The track parameter, $\sigma$, gives an indication of the diffusion 
present for each event.
Figure~\ref{fig:diffusion} shows the variance ($\sigma^2$) 
for cosmic tracks at various drift distances in TDR gas at $B = 1$ and 4~T.
The dependence is linear, as expected from diffusion.

An estimate of the diffusion constant, $D$, is given by the square root of the slope and
the defocusing, $\sigma_0$, is given by the square root of the intercept.
When the technique is applied to simulated data, for a broad
range of input diffusion constants, the estimated diffusion constants are
found to be about 10\% too small, when the noise parameter,
described in appendix~\ref{xyTrackFit}, 
$p_\mathrm{noise}$ is set to 0.01. 
Furthermore, the discrepancy increases for larger noise parameter values.
The simulation, however, appears to describe the effect, since the ratio of diffusion
constants in data to those in the simulated data is found to be constant
within 3\% for 
$0.001 < p_\mathrm{noise} < 0.05$.
Table~\ref{tab:gasprop} shows the results for 
diffusion and defocusing as measured by the
cosmic ray data collected in 2004, compared to the expectation.
The measured diffusion constants have been corrected by scaling by a factor of $1.11\pm0.03$.

The observed drift velocity and diffusion constants are in relatively good
agreement, apart from the diffusion at zero magnetic field.
The defocusing term is larger in the data as compared to the simulation, for
all data sets.

\begin{figure}
\centering
\includegraphics[width=66mm,clip]{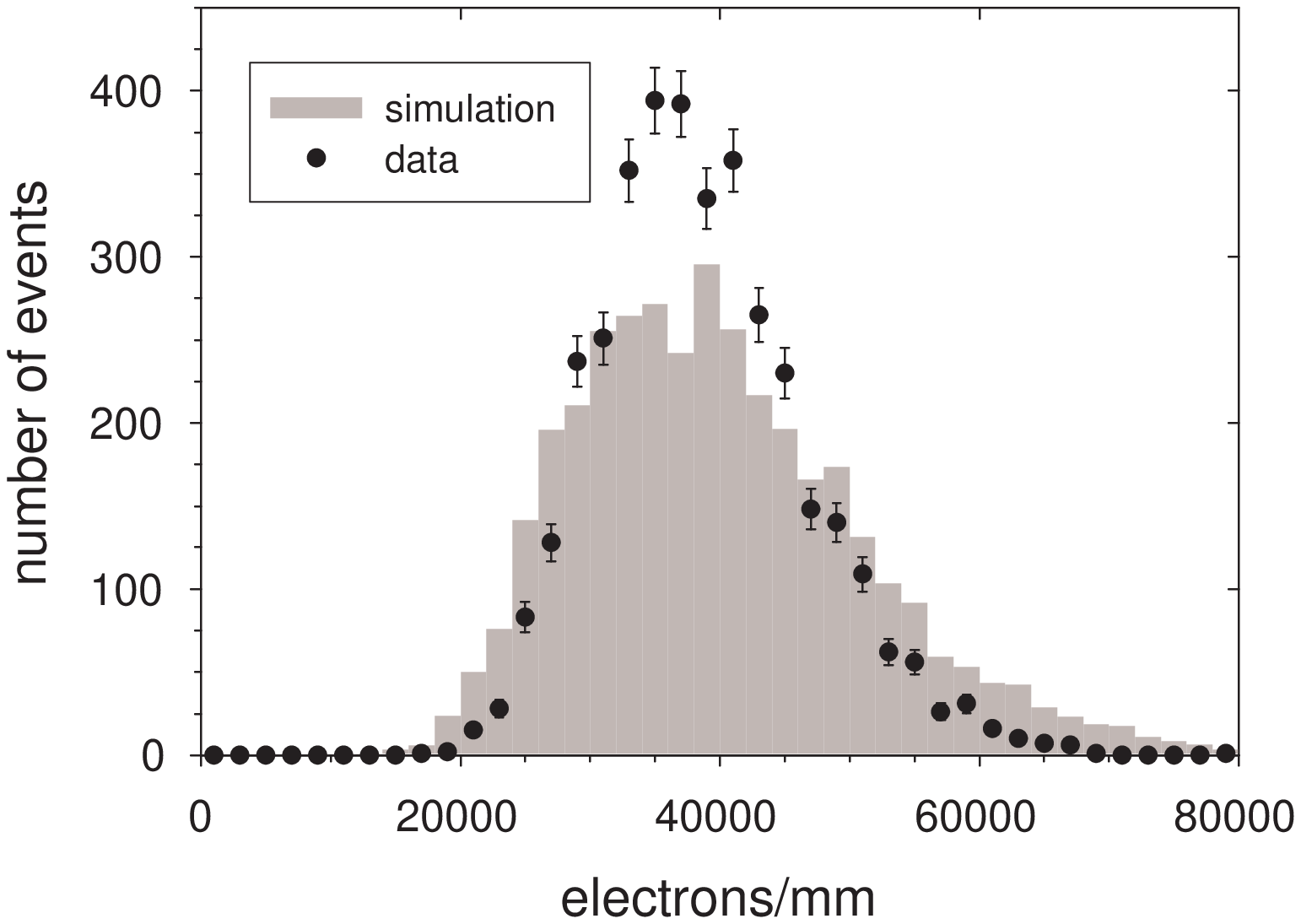} \ \ \ 
\includegraphics[width=66mm,clip]{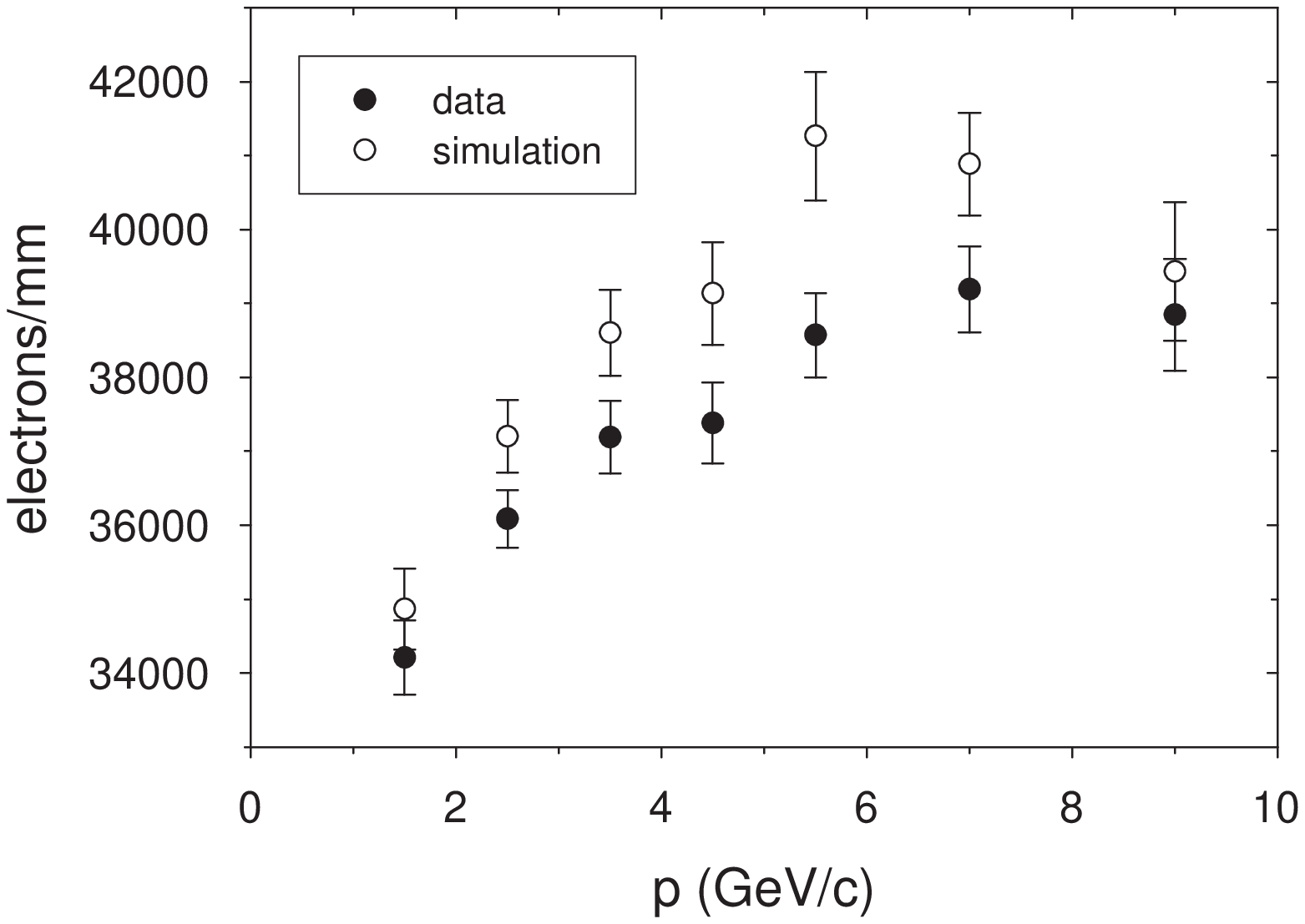}
\caption{
The left figure shows the $dE/dx$ distribution in terms of 
the truncated average number of electrons collected on a pad row per mm
of path length sampled by the row. 
The points show the data from the p5B4w sample 
where the GEM system gain is estimated
to be approximately $8\times10^3$ for this run.
The shaded histogram shows the result from the simulation.
The right figure shows the  mean $dE/dx$ plotted in bins of momentum
for data and simulation.
} \label{fig:dedx}
\vskip 1cm
\end{figure}

The energy loss of cosmic muons traversing the gas is measured using all 11 pad rows, 
including the three very wide pads.
Shown in Fig.~\ref{fig:dedx} is the truncated mean number of electrons collected
on a pad row per mm of path length sampled by the row,
where
the single largest and smallest values are
discarded in calculating the truncated mean.
It is seen that the GEANT3 simulation gives a somewhat broader 
distribution than observed in the data.
The increase in $dE/dx$ as a function of estimated momentum, as deduced by
the track curvature, is seen, as expected.
Overall, the $dE/dx$ resolution is approximately 17\%.
In comparison, an empirical relation~\cite{Walenta:1979bx}
that reproduces the $dE/dx$ performance of many
large scale experiments predicts 16\% for this setup. 
The modest resolution is a result of the short overall sampling length of 86~mm.

\subsection{Track resolution in the pad plane}
\label{trackResolution}
The likelihood track fit procedure
does not involve individual space points along the length of the
track, so there is no possibility to use the scatter of such points
to estimate the transverse resolution per point.
Instead, the resolution is deduced by comparing the transverse coordinate
estimates from fits using different sets of rows.
First, the likelihood fit to all 8 rows of data is used to define a reference.
Next, the data from a single row is used to estimate the horizontal coordinate, 
in a fit that leaves the other track parameters fixed according to the reference.
This is done for all events in a data set and
the distribution of residuals, between the reference and the single row fit,
is fit to a single Gaussian to determine the standard deviation, $\sigma_\mathrm{w}$.
In general, the residual distributions are found to be well described by a single Gaussian.
Because the data from the single row was used in the reference, $\sigma_\mathrm{w}$ will
underestimate the true resolution.
To overcome this, the process is repeated, but
this time the reference fit does not use the information from the single row.
The corresponding standard deviation, $\sigma_\mathrm{wo}$, is an
overestimate of the true resolution because the reference track is not
perfectly measured itself.
To properly estimate the resolution from a single row, 
the geometric mean of these two standard
deviations is used.
This approach is exact for a traditional least squares 
fit~\cite{Carnegie:2004cu}, and is also found to work
well in simulations and with the laser data, as described in Appendix~\ref{lasercheck}.

\begin{figure}
\centering
\includegraphics[width=66mm,clip]{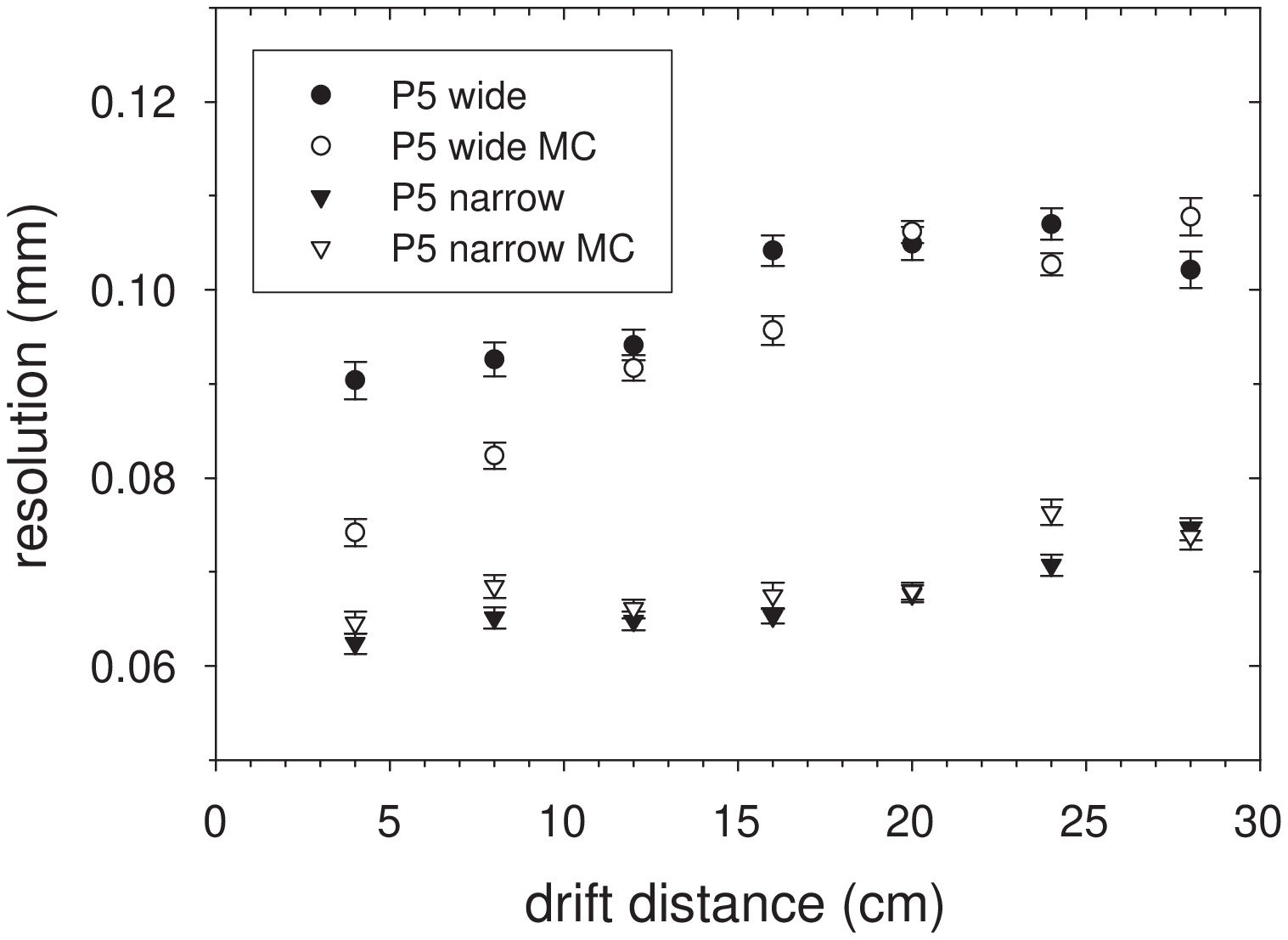} \ \ \ 
\includegraphics[width=66mm,clip]{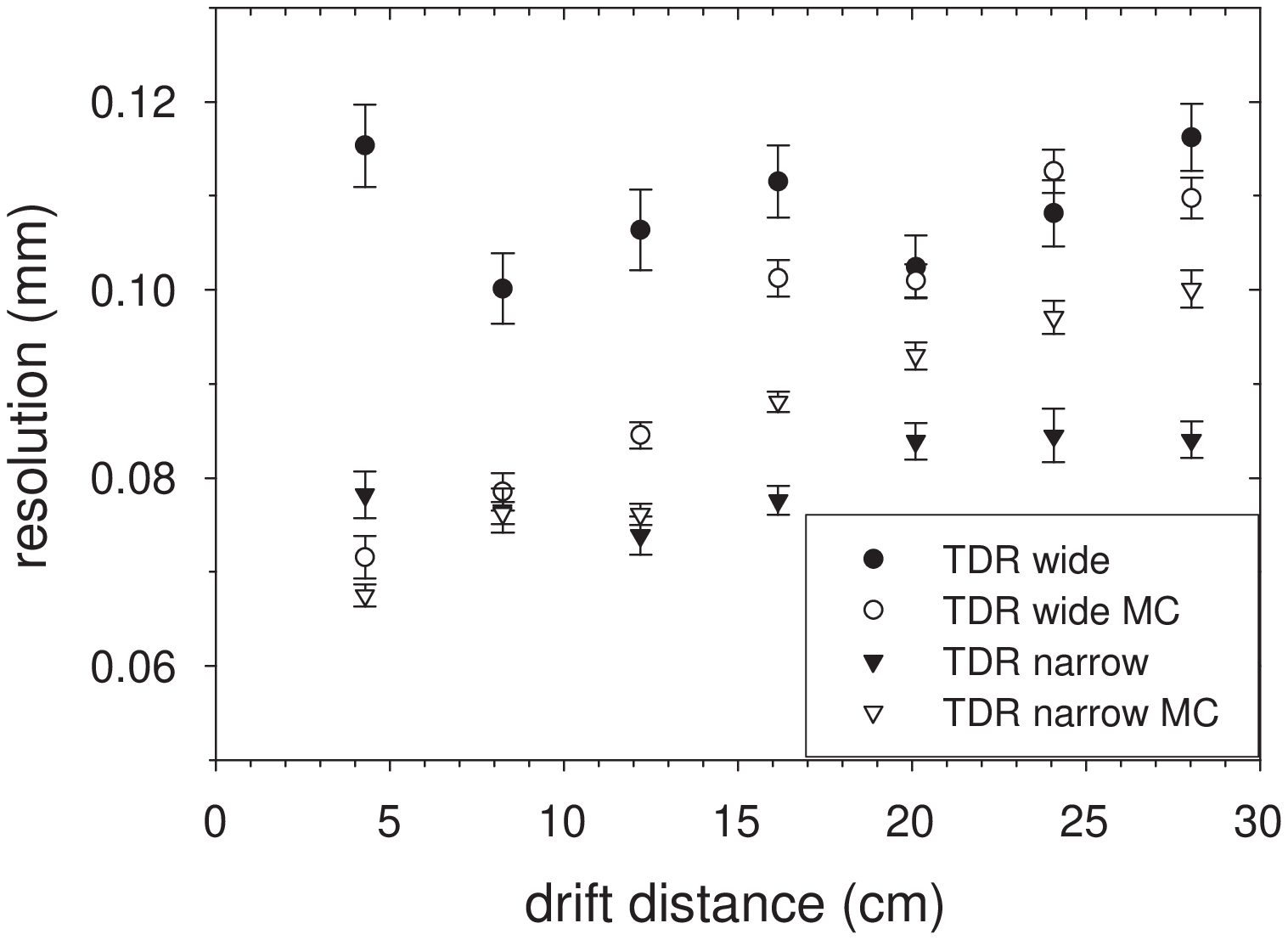}
\caption{
The spatial resolution in the pad plane for a row of
pads with 7~mm height is shown for P5 (left)
and TDR (right) gases as a function of drift distance.
Wide (2~mm) and narrow (1.2~mm) pitch pads are compared, along with results from
simulated data sets (open points, labelled MC).
} \label{fig:resol}
\vskip 1cm
\end{figure}

The single row resolution for cosmic ray tracks measured in a 4~T magnetic
field by the TPC is shown in Fig.~\ref{fig:resol} as a function of drift distance
and compared to the results from simulated data sets.
For both gases, the resolution is improved substantially by reducing the pad
pitch from 2~mm to 1.2~mm, indicating that the limited charge sharing for
the wide pads is a significant factor in the resolution for those pads. 
Likewise, better resolution is seen in P5 gas than TDR gas, which would
be expected, given that the defocusing is larger in the P5 gas, thereby
providing more charge sharing.
The overall resolutions, by combining events from all drift distances are summarized
in table~\ref{tab:resol}.

\begin{table}
\begin{center}
\caption{
Overall single row spatial resolution in the pad plane for the 4 data
sets collected at 4~T and the corresponding simulated data sets.
\vskip 5mm
}
\begin{tabular}{|l|c|c|c|}
\hline
dataset & Resolution      & Resolution      \\[-8pt]
        & [$\mu$m] (data) & [$\mu$m] (sim.) \\
\hline
p5B4w   & $108\pm1$   & $92\pm1$    \\
\hline
p5B4n   & $68\pm1$    & $68\pm1$    \\
\hline
tdrB4w  & $117\pm2$   & $100\pm1$   \\
\hline
tdrB4n  & $83\pm1$    & $87\pm1$    \\
\hline
\end{tabular}
\label{tab:resol}
\end{center}
\vskip 1cm
\end{table}

\begin{figure}
\centering
\includegraphics[width=90mm,clip]{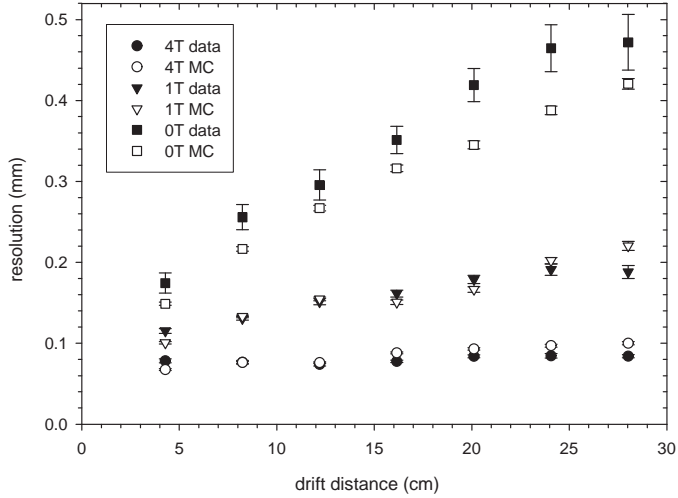} 
\caption{
The spatial resolution in the pad plane for a row of
pads with 7~mm height is shown for different
magnetic field strengths, for the TDR gas with narrow pads.
} \label{fig:resolbfield}
\vskip 1cm
\end{figure}

The resolution is substantially poorer at lower magnetic fields as shown in
Fig.~\ref{fig:resolbfield}.
A stronger dependence on drift distance is seen as expected, due
to the increased transverse diffusion constant for the gas at lower fields.

\begin{figure}
\centering
\includegraphics[width=90mm,clip]{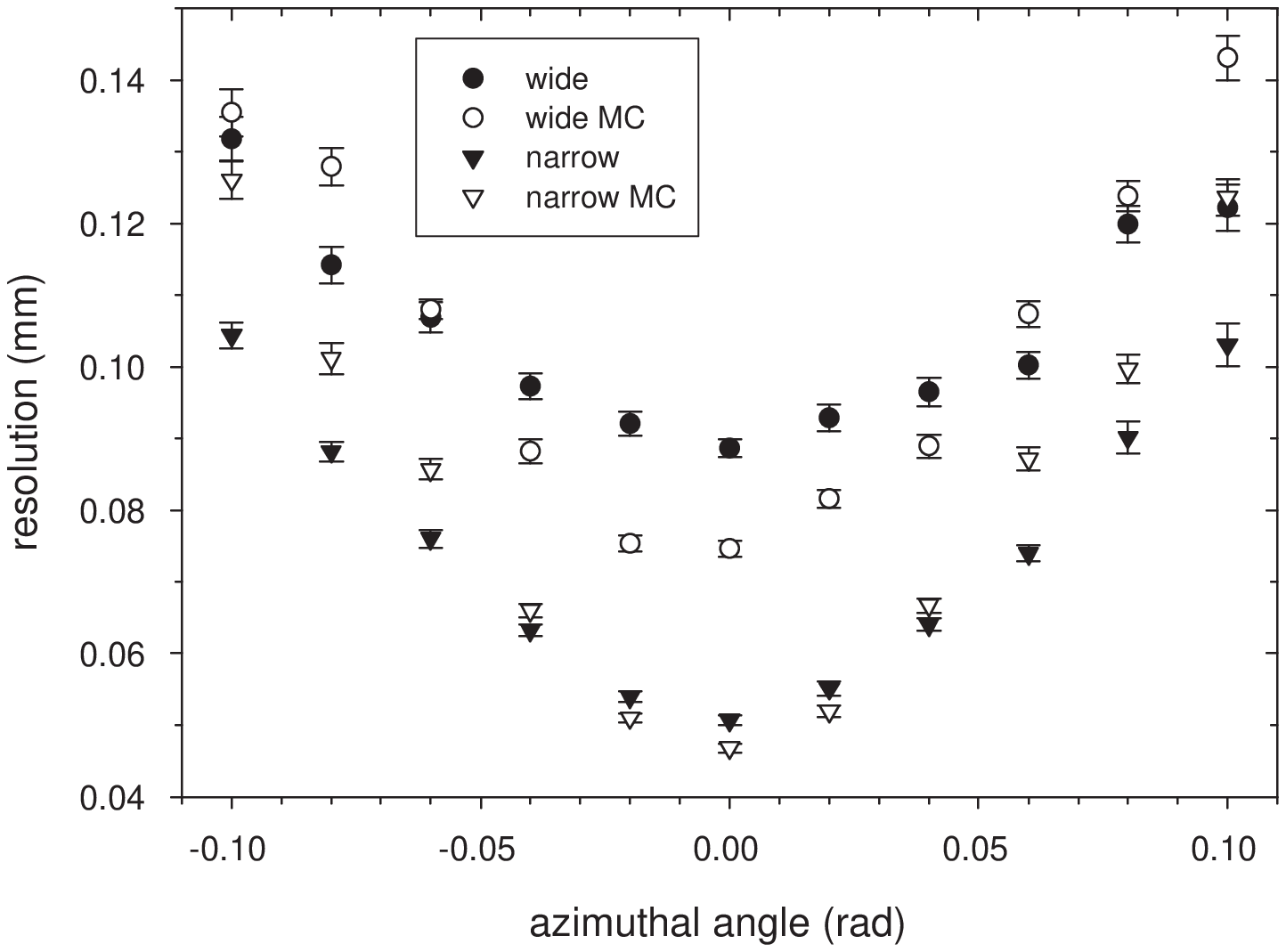} 
\caption{
The spatial resolution in the pad plane for a row of
pads with 7~mm height is shown for different azimuthal
angles is shown for P5 data samples at 4~T.
} \label{fig:phiresol}
\vskip 1cm
\end{figure}

Figure~\ref{fig:phiresol} shows that
the resolution degrades with increasing azimuthal angle, as expected from the
non-uniform ionization of cosmic rays.~\cite{Carnegie:2004cu}
The dependence is stronger for the narrow pads than wide pads.
The simulation has a stronger dependence than seen in the data, which could be related to the
fact that the ionization fluctuations appear to be larger in the simulation sample,
as shown in Fig.~\ref{fig:dedx}.

\begin{figure}
\centering
\includegraphics[width=66mm,clip]{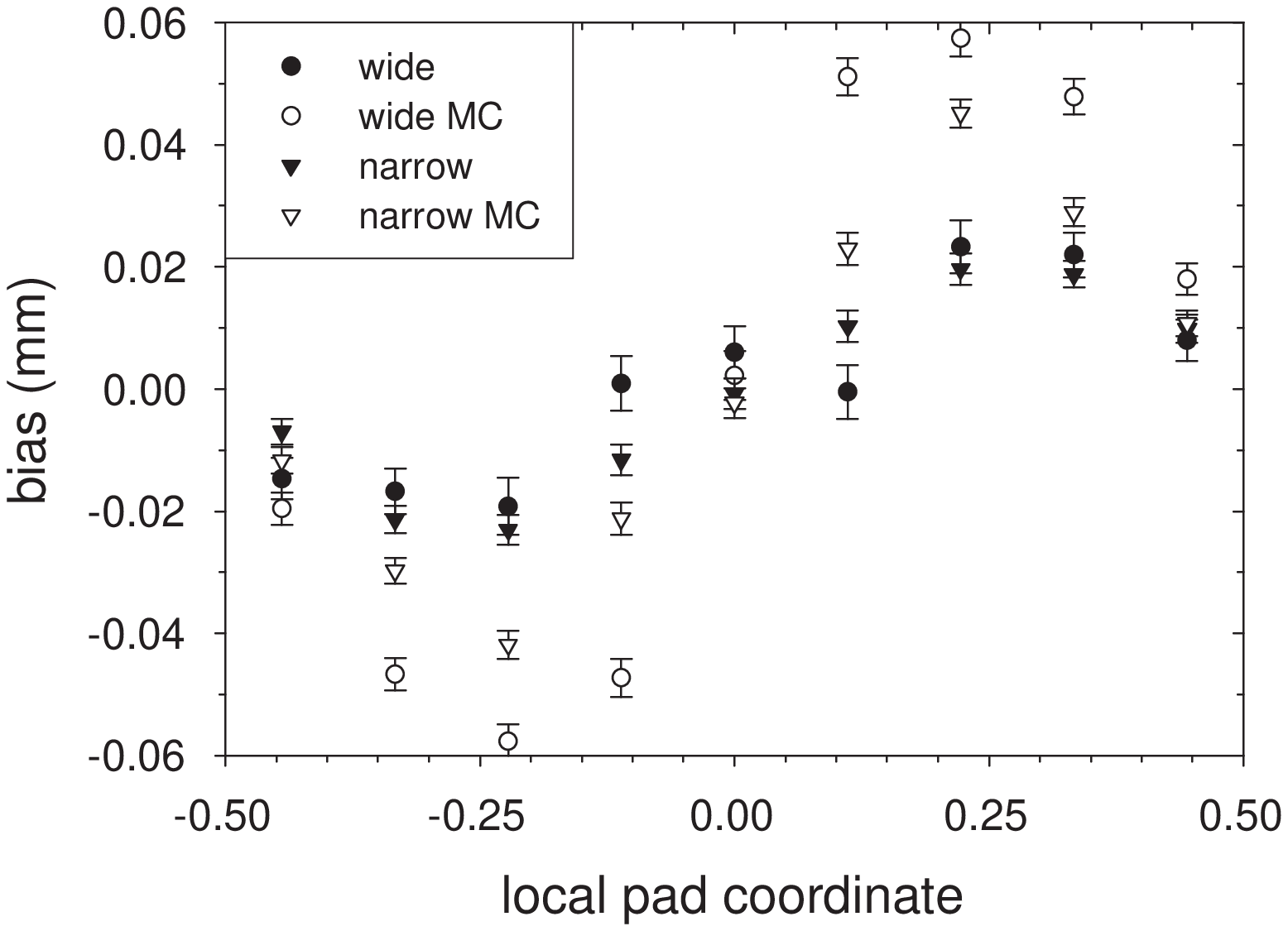} \ \ \ 
\includegraphics[width=66mm,clip]{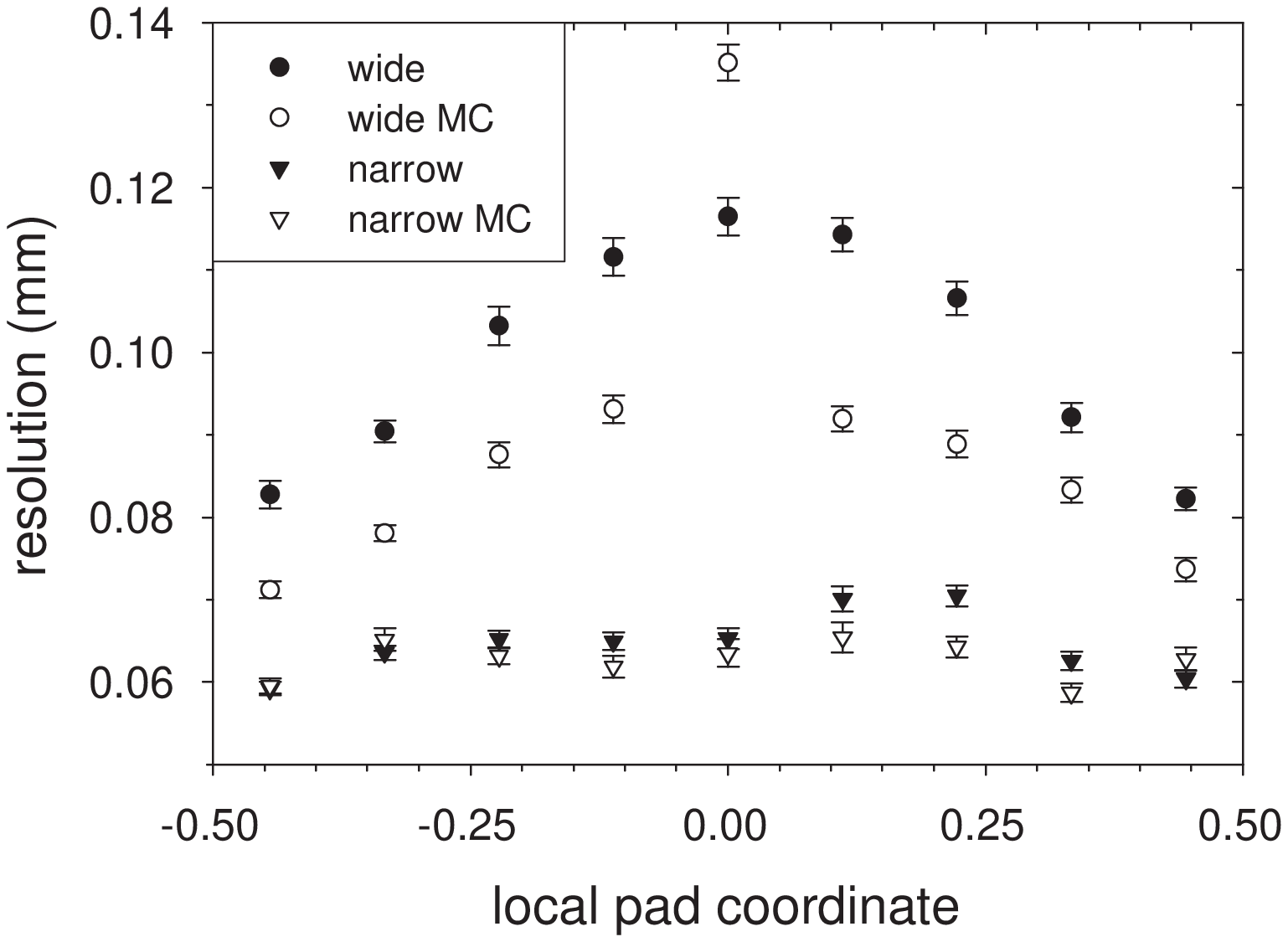}
\caption{
The single row bias and resolution for P5 data at 4T
is shown as a function of local pad coordinate (the location
of the reference track on the pad, mapped to $-0.5$ - $0.5$, where the centre of the pad
is located at 0). The left figure shows a bias dependence that arises from an underestimate
of the charge width, $\sigma$. The right figure shows that for the wide pads, the resolution 
is better near the pad edges.
} \label{fig:biasresb}
\vskip 1cm
\end{figure}

The dependence of the location that a track crosses a pad on the bias and resolution for
a pad row, is shown in Fig.~\ref{fig:biasresb}.
In the figures, the local pad coordinate is the location of the
reference track (fit from all rows) as it crosses the pad, represented as a fraction
of the total pad width. 
A negative bias is seen for tracks with a somewhat negative local pad coordinate and a positive
bias is seen for tracks with a somewhat positive local pad coordinate.
This is a consequence of the fact that the width of the charge cloud, $\sigma$, is underestimated
because of the use of the noise parameter, $p_\mathrm{noise}$, as described in 
section~\ref{generalProperties}.
For smaller values of $p_\mathrm{noise}$, $\sigma$ increases and the amplitude of the 
bias oscillation decreases. 
The magnitude of the bias oscillation is larger in the simulated samples.
In all cases, the bias is smallest at the edges or the centre of the pad, as expected, since
$\sigma$ does not influence the reconstructed transverse coordinate at those locations.
Overall, the effect is small enough that it does not significantly contribute to the overall 
pad row resolution.

For the wide pads, the resolution is best for tracks that cross near the edge of the
pad.
This dependence is expected and described in Appendix~\ref{chargeSharing}.
The simulation shows the same trend, but shows a much more peaked behaviour at the
centre of the pad.
For the narrow pads, the resolution is roughly constant across the pad for data
and in the simulation.

\subsection{Two track resolution in the pad plane}
\label{twoTrackResolution}
\begin{figure}
\centering
\includegraphics[width=90mm,clip]{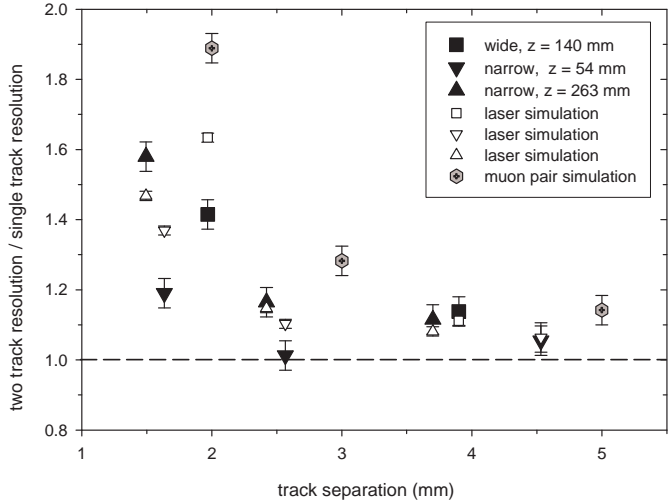} 
\caption{
The degradation in transverse resolution is shown as a function of the transverse
separation between two parallel tracks, for P5 gas at 4T.
The z values shown are the drift distances of the tracks.
The laser data and simulations (open symbols with the same shape) 
show rough agreement. The simulation result
for muons with 2~mm wide pads, 
indicate good resolving power for separations above around 3~mm.
} \label{fig:twotrack}
\vskip 1cm
\end{figure}

To study the capability of the detector to extract track coordinate information for
nearby tracks, two parallel laser beams were brought close together at the same
drift distance.
This represents the most difficult situation for resolving two tracks in a TPC.
By using the beam blockers, events were recorded with individual beams, as well as events
with both beams present.
As described in section~\ref{twoTrackFitting}, the likelihood track fitter was modified
to fit two tracks for this study.
The fractional increase in the standard deviation of the $x_0$ coordinates when both
beams are present is used to quantify the two track separation power.
A simulation of the laser tracks was also performed, where the laser ionization
was assumed to be Poisson distributed.
Figure~\ref{fig:twotrack} shows the degradation in the resolution as the beams are
brought together for the data and simulation, in wide and narrow
pads in P5 gas at 4T.
There is rough agreement with the data and simulation.
To estimate the two track resolution capability for minimum ionizing tracks in the device,
a simulated data set of muon pairs at different separations was generated for the wide pads,
and the results are included in Fig.~\ref{fig:twotrack}.
The track fitter fails to converge for 1\% (5\%) of the muon pair events
for track separations of 3~mm (2~mm).
In summary, this study indicates that for a GEM TPC with 2~mm wide pads,
significant information can be extracted from
a pad row when two tracks are as near as 3~mm in that row.

\subsection{Track resolution in the drift direction}
\label{zTrackResolution}
The focus of this study is on the resolution performance in the pad plane for which the
demands are the greatest and the magnetic field has a direct impact.
For completeness, the observed resolutions in the drift direction are reported in this section.
A linear track fit is performed using the vertical coordinates of the pad rows and the
drift time for the cluster observed on the row.
The residuals of the drift time from each pad row with respect to the full track fit 
are fit to Gaussians to determine the standard deviations.
As before, the geometric mean of the standard deviations determined when
the track fit includes or excludes the pad row is taken to be the resolution.
The resolution is seen to degrade with drift distance, as expected from the
longitudinal diffusion.
The data resolution is somewhat poorer than the values found from the simulation,
as shown in table~\ref{tab:zresol}.
Although the TDR mixture has a lower longitudinal diffusion constant than P5, it has
a faster drift velocity, and with the 50~ns sampling, the resolutions of the two
gases are found to be quite similar.

\begin{table}
\begin{center}
\caption{
Overall single row spatial resolution in the drift direction for the 2004 data
sets and the corresponding simulated data sets.
\vskip 5mm
}
\begin{tabular}{|l|c|c|}
\hline
dataset & Resolution      & Resolution      \\[-8pt]
        & [mm] (data)     & [mm] (sim.) \\
\hline
p5B4w   & $0.99\pm0.01$    & $0.83\pm0.01$    \\
\hline
p5B4n   & $1.02\pm0.01$    & $0.81\pm0.01$    \\
\hline
tdrB4w  & $0.99\pm0.01$    & $0.85\pm0.01$   \\
\hline
tdrB4n  & $0.95\pm0.01$    & $0.87\pm0.01$    \\
\hline
tdrB1n  & $0.99\pm0.01$    & $0.90\pm0.01$    \\
\hline
tdrB0n  & $1.15\pm0.02$    & $0.93\pm0.01$    \\
\hline
\end{tabular}
\label{tab:zresol}
\end{center}
\vskip 1cm
\end{table}

\section{Conclusions}
\label{concl}
The results presented in this paper indicate that a
GEM TPC with pad readout is a viable candidate for the central
tracker for an experiment at the International Linear Collider.
When operated in a 4~T magnetic field with modest size pads, 
a full size TPC could collect
some 200 pad-row measurements each with transverse spatial resolution
of approximately 100~$\mu$m,
which should be sufficient to achieve the transverse momentum resolution
goal of
$\sigma(1/p_t) = 2 \times 10^{-4}$~(GeV/c)$^{-1}$.~\cite{Behnke:2001qq}
In addition, the resolution does not degrade significantly
for nearby tracks, provided their separation is more than about 1.5~times
the pad width.

The relatively simple simulation package reproduces many of the general features
seen in the data.
It should be useful, therefore, in optimizing
the configuration of GEM TPCs for future experiments at the ILC and elsewhere.

\section{Acknowledgements}
This work would not have been possible without the assistance from the DESY
laboratory in providing the superconducting magnet test facility and the UV laser, and
we especially thank Thorsten Lux and Peter Wienemann
for assistance at the laboratory.
We would like to thank Michael Ronan for providing the STAR TPC electronics
used in this work.
The work of Vance Strickland in designing and constructing the TPC and
Mark Lenckowski in designing and constructing the laser delivery system was invaluable.
We were assisted by undergraduate students, Brie Hoffman, Camille Belanger-Champagne,
and Chris Nell,
and by technical staff at the University of Victoria and TRIUMF.
Our work benefited from collaboration with the TPC group at
Carleton University and the worldwide LC-TPC group.
This work was supported by a grant by the Natural Sciences and Engineering
Research Council of Canada.

\appendix

\section{Resolution and Charge Sharing in a GEM TPC}
\label{chargeSharing}
Consider a charge cloud of $n_p$ electrons arriving at the GEM plane.
Due to diffusion in the drift volume, the electrons are distributed in the
transverse direction, $x$, with standard deviation $\sigma_d$.
If the GEMs provide a gain $g$ (with negligible variance), the defocusing
adds a variance $\sigma_0^2$ to the $x$ distribution, and
if the $x$ coordinate of all of the resulting electrons are measured with
no uncertainty, then the variance of the mean $x$ coordinate is
\begin{equation}
\sigma_{\bar{x}}^2 = 
     {1\over n_p}\left( \sigma_d^2 + {\sigma_0^2\over g} \right) .
\end{equation}

To achieve the diffusion limit, the GEM term (the second term) must be 
much smaller than the diffusion term.
In reality, the GEM term has additional contributions from
the variance in the gain, and the uncertainties in measuring
the $x$ coordinates for each electron.

In a GEM-TPC the $x$ coordinate for each electron is not measured, but
rather the charge shared by neighbouring pads is used to deduce the
mean $x$ coordinate.
The degradation in resolution due to using large pads can be
understood analytically.
Consider two neighbouring semi-infinite pads with the boundary at
$x=0$.
If the electrons are distributed according to the pdf $G(x)$, the
expectation for the fraction of electrons over the positive-$x$ pad is
\begin{equation}
\left< F\right> = \int_0^\infty G(x)\,dx \ \ .
\end{equation}
If $G(x)$ is Gaussian, with mean $\mu$, standard deviation $\sigma$, the
estimate $\hat{\mu}$ determined from the observed fraction $F$ has
variance,
\begin{equation}
\sigma_{\hat{\mu}}^2 = 2 \pi \sigma^2 e^{\,\mu^2/\sigma^2} \sigma_F^2
\end{equation}
where the variance of $F$ is binomial,
$
\sigma_F^2 = \left< F\right> \left( 1 - \left< F\right> \right) / n
$
and $n$ is the number of electrons.
If $\mu = 0$, so that the mean of the pdf is at the border between
the pads, the variance on the estimate $\hat{\mu}$ is
$
\sigma_{\hat{\mu}}^2 \approx 1.6 \sigma^2/n
$
or in other words a factor 1.6 larger than the variance that would
result in perfect $x$ coordinate measurements for each electron.
As the mean of the pdf moves away from the pad boundary, the variance gets larger;
for $\mu = 1 \sigma$ ($2 \sigma$) the factor increases to 2.3 (9.0).
To keep the GEM contribution to the resolution small, the pad width
needs to be less than about 3-4 $\sigma$, so that
$\mu$ is less than 1.5-2 $\sigma$.
The effects of noise and thresholds may limit the pad sizes even further.

\section{Maximum Likelihood Track Fit in the Pad Plane}
\label{xyTrackFit}
The maximum likelihood track fit uses
a simple model to describe the way electrons are spread about the readout pads.
Due to diffusion, the electrons are assumed to be distributed in a 
guassian fashion about the projection of the particle trajectory onto the readout 
pad plane.
The resulting ``Line-Gaussian'' density function is a convolution
of the ``trajectory density function'' and a two dimensional 
isotropic Gaussian probability density function.
The standard deviation of the Gaussian, $\sigma$, is a
free parameter in the fit, like the other track parameters.
The fitted value provides an estimate of the diffusion for each event.
For simplicity in the analysis, $\sigma$ is assumed to be constant along the
the length of the track,
a good approximation for the data samples described in this paper.

The implementation of this model assumes that rectangular pads of equal height
are arranged in a parallel fashion into horizontal rows and that tracks
cross several rows.
The height of the row is assumed to be much smaller than the radius of
curvature, so that the tracks can be considered to be straight within the vertical
extent of the row.

The true trajectory density function (tdf) for a typical track is non-uniform along its
path due to ionization fluctuations.
The model, however, assumes that within a row, the tdf is uniform.
The likelihood, $\mathcal L_\mathrm{row}$,
of observing the distribution of charge is calculated for each row,
and the product of the likelihoods for all the rows defines the overall 
likelihood function, $\mathcal L$.
In this way, the model implicitly allows the tdf to vary from row to row.

The expected charge collected by a rectangular pad is determined by integrating a
uniform Line-Gaussian density function over the physical region of the pad and
is proportional to:
\begin{equation}
\begin{array}{rcl}
I(b,\phi,\sigma,h,w) &=& \displaystyle{\int_{-w/2}^{w/2}\int_{-h/2}^{h/2}}
{dx\,dy\over\sqrt{2\pi}\sigma} 
\exp\left(-{[(x-b)\cos\phi+y\sin\phi]^2\over 2\sigma^2}\right) \\
& & \\
&=& \eta(b,\phi,\sigma,h,w) - \eta(b,\phi,\sigma,-h,w) \\
& &+ \eta(b,\phi,\sigma,-h,-w) - \eta(b,\phi,\sigma,h,-w) \\
& & \\
\eta(b,\phi,\sigma,h,w) &=& \displaystyle{{1\over\cos\phi\sin\phi}
     \xi\left(\left(b+{w\over2}\right)\cos\phi+{h\over2}\sin\phi,\sigma\right)} \\
& & \\
\xi(u,\sigma) &=& \displaystyle{{u\over2}{\rm erf}\left({u\over\sqrt{2}\sigma}\right)
                 + {\sigma\over\sqrt{2\pi}}\exp\left({-u^2\over2\sigma^2}\right)}\\
\end{array}
\end{equation}
where $b$ is the horizontal distance between the pad centre and the track,
$\phi$ is the local azimuthal angle of the straight line segment in the pad row,
and $h$ and $w$ are the height and width of the pads.
The local azimuthal angle for a row centred at $y = y_\mathrm{row}$ is
\begin{equation}
\sin\phi = \sin\phi_0 - y_\mathrm{row}/r \ .
\end{equation}
The horizontal distance to the track from the pad centred at $x = x_\mathrm{pad}$ in
that row, can be expressed in a power series of $1/r$ as
\begin{equation}
b = x_0 - x_\mathrm{pad} - y_\mathrm{row}\tan\phi_0
+ {\textstyle\frac{1}{2}} y^2_\mathrm{row}\sec^3\phi_0{1\over r}
- {\textstyle\frac{1}{2}} y^3_\mathrm{row}\tan\phi_0\sec^4\phi_0{1\over r^2} \ .
\end{equation}

In order to calculate the likelihood of the observed charge distribution 
in a single row, an approximate model is used.
The model is developed on the basis that signals in a row result from a relatively small
number, $n \approx O(100)$, of electrons liberated in the 
ionization process that are subsequently amplified in the GEM structure.
As an approximation, the
distribution of charge is assumed to be multinomial with $n$ observations.
This approximation is exact in
the limit that the diffusion within and after the GEM structure is
small compared to the diffusion in the drift region,
the fluctuations due to electronics noise is small compared to that due to
primary electron statistics, and that thresholds are small.
This model was initially developed to study data from tests without magnetic fields, 
in which the dominant
contribution to diffusion was in the drift volume.
For certain gases and operating conditions in strong magnetic fields, 
the situation is reversed, whereby the diffusion primarily arises in the GEM structure.
In fact this feature is desirable in order to ensure that there is good charge
sharing across relatively large pads, as described in section~\ref{intro}.
Nevertheless, the model performs very well in either circumstance.

If $N_i$ electrons are collected by pad $i$ for a system with total GEM gain, $g$,
the equivalent number of the original ionization electrons associated 
to the pad is simply defined as
\begin{equation}
n_i = N_i / g \ .
\end{equation} 
These values are non-integer, but a direct continuum extension of the
multinomial distribution, yields the log likelihood function,
\begin{equation}
\log{\mathcal L}_\mathrm{row} = \sum_i n_i \log p_i + \mathrm{constant}
\end{equation}
where $p_i$ is the probability for a primary electron from the track
to be associated to pad $i$,
\begin{equation}
p_i = I(b_i,\phi_i,\sigma,h_i,w_i) / \sum_j I(b_j,\phi_j,\sigma,h_j,w_j)
\end{equation}
where $j$ runs over all pads in the row containing pad $i$.

A spurious signal in a pad far from the track can cause problems for
the likelihood calculation, since the calculated probability for electrons
to be present there can be vanishingly small.
To make the track fit robust to spurious signals, the
probability for a primary electron to be associated with a pad is modified by adding
a small constant, $p_\mathrm{noise}$,
\begin{equation}
p_i \rightarrow \frac{p_i + p_\mathrm{noise}}{1 + n_\mathrm{pad}\,p_\mathrm{noise}}  
\end{equation}
where $n_\mathrm{pad}$ is the number of pads in the row.
The fitted track parameter values, apart from $\sigma$, 
are found not to strongly depend on
the choice of $p_\mathrm{noise}$, for the data analyzed in this paper, where
$p_\mathrm{noise} \approx 0.01$.

\section{Study of resolution determination method}
\label{lasercheck}
The method used in this paper to form the estimated single row resolution (ESSR)
has been tested using simulated samples and laser data.

For simulated samples, the true values of the track parameters are known.
Events were simulated with the ionization 
following a straight trajectory, and the
resulting pad signals were fit to straight tracks (fixing $1/r = 0$). 
The resolution is directly determined from the standard deviation of the
residuals of the fit $x_0$ values with respect to the true values.
The direct single row resolution is defined to be $\sqrt{8}$ times this
standard deviation, since 8 rows are used in the fit.
The ESRR is found to agree with the direct result to within 10\%.
When the same data is fit to curved tracks (allowing $1/r$ to float), the
ESRR is unaffected, but the standard deviation of the residuals 
for the direct result increases
due to the correlation between $x_0$ and $1/r$.

Laser data was used in a similar way to check the ESRR.
It was found that the laser system was quite steady;
the mean of the $x_0$ from track fits to the laser data
typically drifted by less than 10~$\mu$m over a period of 12 hours.
In the data with two laser beams present at large drift separations, 
however, the reconstructed $x_0$ values for the two tracks were found to be
correlated.
This is interpretted as due to an angular jitter of the laser and
a simple model is used to describe this situation.
The measured track parameter $x_0$ for track $i$ ($i=1$ or 2 - the near or far track),
is described as an outcome of a random variable $X_i$,
\begin{equation}
X_i = \mu_i + R_\ell + R_i
\end{equation}
where $R_\ell$ is a random variable of mean zero describing the jitter of the laser
system and $R_i$ is a random variable of mean zero describing 
the random nature of the $x_0$ measurement by the TPC.
The covariance, $\left<(X_1-\mu_1)(X_2-\mu_2)\right>$, isolates the contribution from
the laser jitter, and its square root is found to be about 10~$\mu$m, corresponding
to an angular jitter of about 5~$\mu$rad.

Only data with two laser pulses are used, so that the laser jitter term can
be removed, and the standard deviation of the remainder is multiplied by
$\sqrt{8}$ to form the direct single row resolution.
The direct single row resolution is determined from fits with $1/r = 0$, in order
that the correlation between $x_0$ and $1/r$ not influence the resolution measurement.

\begin{figure}
\centering
\includegraphics[width=90mm,clip]{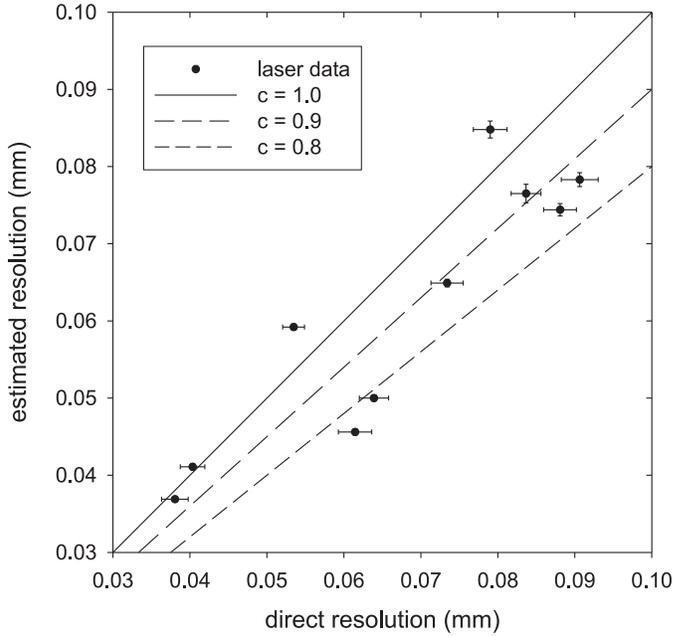} 
\caption{
The estimated single row resolution is compared to a direct resolution estimate
from laser data collected with narrow pads and P5 gas at 4T.
The error bars indicate the statistical uncertainties of the resolution estimates.
In each case, the laser jitter contribution has been removed in the
calculation of the direct resolution.
The lines correspond to ESRR = c $\times$ direct.
} \label{fig:lasercheck}
\vskip 1cm
\end{figure}

This is compared to the ESRR for the same data sets
in Fig.~\ref{fig:lasercheck}.
The fits used in the ESRR calculation allowed $1/r$ to float, as is done when
estimating the resolution with the cosmic ray data.
The ESRR is found to be somewhat smaller than the
direct resolution; it appears that ESRR  $\approx (0.9\pm0.1) \times$ direct.

In summary, the simulation and laser studies show that the method used in this
paper to estimate the single row resolution of a GEM TPC appears to be
correct within a systematic uncertainty of about 10\%.

\end{document}